\documentstyle[12pt,titlepage]{article}
\begin{document}
\begin{titlepage}
\title{
Thermodynamic Length, Time, Speed and Optimum Path
to Minimize Entropy Production}
\author{L. Di\'osi\thanks{E-mail: diosi@rmki.kfki.hu}\\
KFKI Research Institute for Particle and Nuclear Physics\\
H-1525 Budapest 114, P.O.Box 49, Hungary\\
	K. Kulacsy\thanks{E-mail: H10575kul@huella.bitnet}\\
KFKI Atomic Energy Research Institute\\
	B. Luk\'acs\thanks{E-mail: lukacs@rmki.kfki.hu}\\
KFKI Research Institute for Particle and Nuclear Physics\\
	A. R\'acz\thanks{E-mail: racz@fserv.kfki.hu}\\
KFKI Atomic Energy Research Institute\\
{\it e-archive ref.: cond-mat/9503174}
}
\date{July 28, 1995}
\maketitle
\begin{abstract}
In addition to the Riemannian metricization of the thermodynamic state
space, local relaxation times offer a natural time scale, too. Generalizing
existing proposals, we relate {\it thermodynamic} time scale to the standard
kinetic coefficients of irreversible thermodynamics. Criteria for minimum
entropy production in slow, slightly irreversible processes are discussed.
Euler-Lagrange equations are derived for optimum thermodynamic control for
fixed clock-time period as well as for fixed {\it thermodynamic} time
period. Only this latter requires constant thermodynamic speed as the
optimum control proposed earlier. An easy-to-implement stepwise algorithm is
constructed to realize control at constant thermodynamic speed.
Since thermodynamic time is shown to correspond to the number of steps,
thus the sophisticated task of determining thermodynamic time in real
control problems can be substituted by measuring ordinary intensive
variables.  Most remarkably, optimum paths are Riemannian geodesics which
would not be the case had we used ordinary time.

\end{abstract}
\vfill
\end{titlepage}

\section{Introduction}

Standard equations of irreversible thermodynamics have been known for many
decades. Investigations on optimizing finite-time thermodynamic processes
controlled by external reservoirs date from the 80's.
An instructive exposition
of the problem at the time was published in Ref.~\cite{AndSalBer84}.
Basically,
the problem consists of finding the best path in the state space along which
one drives the system from a given equilibrium state to another.

A particular approach to the problem of optimally controlling finite-time
thermodynamic processes takes its origin from the natural geometric
structure of the thermodynamic state space \cite{Wei,Rup,Dio}. It would seem
straightforward to expect that the geodesic path should somehow be related
to the optimum path connecting the given initial and final equilibrium
states.  A few years ago it was pointed out \cite{SalNul8588,SalBer83} that
the natural relaxation time $\tau$ plays a fundamental role in devising
optimum cooling strategies, e.g. in computer simulated annealing. Quite
recently, Andresen and Gordon \cite{AndGor94} have shown that the strategy
of constant thermodynamic speed \cite{SalNul8588} is related to a {\it
certain} minimum of entropy production. In the present work we reconsider
these ideas and make a definite progress.

The problem itself is illustrated by the finite-time cooling process\break
(Sect.~2),
its dynamics is described by a phenomenological cooling-equation. Then,
applying kinetic equations from standard irreversible thermodynamics, we
generalize the concept of thermodynamic time and speed for any number of
control variables (Sect.~3). Furthermore we invoke standard Euler-Lagrange
equations to obtain the finite-time path of minimum entropy production and
we make explicit the role of variational conditions. Most importantly, we
prove that the optimum path is geodesical so a broader generalization of
the old principle of constant thermodynamic speed is achieved. Even an
older belief about the significance of Riemannian metric in thermodynamics
might get justified (Sect.4). Finally, we construct an iterative algorithm
to realize thermodynamic processes at constant thermodynamic speed, also
giving a genuine control-theoretic interpretation of thermodynamic time
itself (Sect.~5).

\section{Example: Finite-time cooling}

Consider a system in thermal contact with a large reservoir and let $T$ and
$T_r$ denote their respective temperatures. In general, both $T$ and $T_r$
will depend on time $t$. At the initial time $t_i$, assume the system and
the reservoir are in equilibrium with each other at temperature $T_i$, i.e.,
\begin{equation}
T_r(t_i)=T(t_i)=T_i.
\end{equation}
Then start to decrease the reservoir's temperature
$T_r$ from $T_i$ to $T_f$,
consuming fixed finite time $t_f-t_i$,
i.e., choose a given function $T_r(t)$
so that $T_r(t_f)=T_f$. The system's temperature $T(t)$ decreases due to
permanent heat transfer to the reservoir and will always be retarded with
respect to the reservoir's current temperature $T_r(t)$ by some positive
$\Delta T(t)\equiv T(t)-T_r(t)$. Throughout this paper, we consider
slightly irreversible processes when, e.g., $T_r(t)$ changes slowly enough
to allow the heat transfer to satisfy Newton's law and the following
equation is expected to drive the permanent relaxation of the system's
temperature $T$:
\begin{equation}
\dot T=-{1\over\tau(T)}\Delta T
\label{coolingeq}\end{equation}
provided the local relaxation time $\tau$ changes little between
$T$ and $T_r$. It will be useful to re-scale the clock-time parameter.
Introducing {\it thermodynamic time} $\xi$ was proposed
earlier in Refs.~\cite{SalNul8588,SalBer83,AndGor94}.
The relation of the two scales relies upon the local relaxation time
$\tau(T(t))$ along the cooling process:
\begin{equation} d\xi=dt/\tau.
\label{dxi}\end{equation}
In the new variable, Eq.~(\ref{coolingeq}) takes a simple form:
\begin{equation}
T^\prime\equiv{dT\over d\xi}=-\Delta T.
\end{equation}
We note that this equation has the following explicit solution with the
initial condition $\xi_i=0$:
\begin{equation}
T(\xi)=e^{-\xi}
\Bigl(T_i
+\int_{0}^{\xi}T_r(\xi^\prime)e^{\xi^\prime} d\xi^\prime\Bigr).
\end{equation}

Now the basic goal is to single out "optimum" cooling paths.
Following e.g. Andresen and Gordon \cite{AndGor94} one requires
that the optimum cooling happen with maximum reversibility, i.e.,
at minimum total entropy production. The entropy production rate of the
cooling process is
\begin{equation}
\dot S=C(T)\dot T\left({1\over T}-{1\over T_r}\right)
\label{dotSdef}\end{equation}
where $C(T)$ is the specific heat of the system. In case of sufficiently
slow cooling this expression reduces to
\begin{equation}
\dot S=C(T)(\dot T/T)^2\tau(T)
\label{dotScool}\end{equation}
where we applied Eq.~(\ref{coolingeq}). The common criterium of optimum is
\begin{equation}
\int_{t_i}^{t_f}\dot S dt = min.
\end{equation}
where the overall time $t_f-t_i$ of the process is fixed. This optimum
is achieved when
\begin{equation}
\dot S={C(T)\over T^2}\dot T^2\tau(T)=const.
\end{equation}
There is, however, a remarkable alternative to this optimum, because one can
choose different boundary conditions. Instead of clock-time $t_f-t_i$
the thermodynamic lapse $\xi_f-\xi_i$ of the cooling can be fixed as well.
Then the optimum cooling becomes different; it will correspond to constant
entropy rate versus thermodynamic time:
\begin{equation}
S^\prime\equiv \tau\dot S={C(T)\over T^2}T^{\prime 2}=const.
\end{equation}
This condition has a challenging geometrical interpretation \cite{AndGor94}.
$S^\prime$ is the
square of the thermodynamic speed of the cooling process:
\begin{equation}
S^\prime=\Vert T^\prime \Vert^2,
\end{equation}
calculated with the entropic metric defined by the quadratic norm
\begin{equation}
\Vert dT \Vert^2\equiv {C(T)\over T^2} (dT)^2.
\label{dTnorm2}\end{equation}
Hence the corresponding optimum process is called cooling at constant
thermodynamic speed (cf. Refs.~\cite{SalNul8588,SalBer83,AndGor94}).
In Sect.~4 we shall prove that the principle of constant thermodynamic
speed also applies for optimum finite-time thermodynamic processes affecting
more (than one) variables. First, in Sect.~3 we must generalize the concept
of thermodynamic length, time, and speed for such finite-time processes.

\section{Thermodynamic length and time}

Let the equilibrium states of a given thermodynamic system be characterized
by the $n+1$ extensive variables; the vector
\hbox{$X\equiv(X^1,X^2,\dots,X^n)$}
will parametrize the manifold of state space while $X^{n+1}$ remains fixed.
Following Refs.~\cite{Wei,Rup,Dio} one defines a metric tensor $g$ on the
manifold of equilibrium states, derived from the entropy $S(X)$:
\begin{equation}
g_{ik}(X)=-{\partial^2S(X)\over\partial X^i \partial X^k}.
\label{gik}\end{equation}
If one took the entropic intensive variables
\hbox{$Y_k=\partial S(X)/\partial X^k;$} \hbox{$k=1,2,\dots,n$}
instead of the extensive variables $X$ then the metric tensor would be the
inverse
\hbox{$g^{-1}\equiv[g^{ik}]$} of \hbox{$g\equiv[g_{ik}]$}. Hence, in obvious
notations, the thermodynamic line-element square takes the following
alternative forms:
\begin{equation}
\Vert dX\Vert^2
=(dX\vert g\vert dX)=\Vert dY\Vert^2=(dY\vert g^{-1}\vert dY).
\label{dl2}\end{equation}
Consider now a certain path \hbox{$\{X(t);t_i\leq t\leq t_f\}$} in the
thermodynamic state space. The {\it thermodynamic length} $\ell$ of the path
takes the (alternative) forms
\begin{equation}
\ell =\int_{t_i}^{t_f} \Vert dX\Vert
     =\int_{t_i}^{t_f} \Vert dY\Vert.
\label{l}\end{equation}
Obviously, the length of a path is independent of the choice of coordinates
and even of the parametrization of the path itself. No intrinsic relation
manifests itself between the thermodynamic length and the clock-time
parameter $t$.

In order to obtain a natural time scale along a given path, we first have to
invoke standard concepts of irreversible thermodynamics. Consider a
reservoir in equilibrium at some state variables $X_r$ and bring it into
contact with the system which is in equilibrium at $X$. Then, the state
of the system will converge to the state of the reservoir.
The standard form of the relaxation equations reads \cite{Lan}:
\begin{equation}
\dot X^i=\gamma^{ik}\Delta Y_k
\label{dotx}\end{equation}
where $\Delta Y=Y_r-Y$ is the deviation from the equilibrium in terms of the
entropic intensive variables, and $\gamma=[\gamma^{ik}]$ is the matrix of
kinetic coefficients; it is symmetric and positive. During this process of
relaxation, entropy $S$ will be produced at rate
\begin{equation}
\dot S  =(\dot X\vert\Delta Y).
\label{dots}\end{equation}
If $\Delta Y$ is small, one can write
\begin{equation}
\Delta Y=-g\Delta X.
\label{Dy}\end{equation}
The relaxation equation (\ref{dotx}) takes the alternative forms:
\begin{equation}
\dot X=-\gamma g\Delta X~~~~~or~~~~~\dot Y=-g\gamma\Delta Y.
\label{dotxy1}\end{equation}
Using the equation (\ref{Dy}) and the equation
\begin{equation}
\dot Y=-g\dot X
\label{dotY}\end{equation}
alternative expressions of entropy production rate (\ref{dots}) follow:
\begin{equation}
\dot S  =(\dot   X\vert \gamma^{-1}      \vert \dot   X)
	=(\dot   Y\vert (g \gamma g)^{-1}\vert \dot   Y).
\label{dots1}\end{equation}

Based on the above standard equations, an effective relaxation time $\tau$,
depending on the path's local direction $\dot X$, can be defined.
Recall from Sect.~2 that the system's path $X(t)$ is driven by a certain
reservoir path $X_r(t)$ and the system's retardation
$\Delta X\equiv X-X_r$ is proportional to the current velocity $\dot X$
of the process:
\begin{equation}
\Delta X=-(\gamma g)^{-1}\dot X
\label{DeltaX}\end{equation}
In fact, this equation is formally identical with the
relaxation equation (\ref{dotxy1}) which remains valid if $X_r$
becomes a slowly varying function of time.
Observe that the longitudinal (i.e. parallel to
$\dot X$) projection of the above equation implies a certain effective
relaxation time $\tau$. Introduce the longitudinal component of the
retardation:
\begin{equation}
\Delta X^\parallel
\equiv{1\over\Vert\dot X\Vert}(\Delta X\vert g\vert \dot X)
\end{equation}
Substituting Eq.~(\ref{DeltaX}) yields the longitudinal relaxation equation
(note that $\dot X^\parallel=\Vert\dot X\Vert$):
\begin{equation}
\Delta X^\parallel=-\tau \Vert\dot X\Vert
\end{equation}
where
\begin{equation}
\tau=-{1\over\Vert\dot X\Vert^2}(\Delta X\vert g\vert \dot X).
\end{equation}
This can be rewritten in the following equivalent form:
\begin{equation}
\tau=-{1\over\Vert\dot X\Vert^2}(\dot X\vert \gamma^{-1}\vert \dot X).
\label{tau2}\end{equation}
Indeed, as it is clearly seen from this form, the effective relaxation time
$\tau$ is positive and only depends on the direction of $\dot X$ but not on
its magnitude (as long as it is moderate).
Eq.~(\ref{tau2}) have the compact form
\begin{equation}
\tau={\dot S\over\Vert\dot X\Vert^2},
\label{tau3}\end{equation}
showing up $\tau$'s invariance if one changes the representation of the
states from extensive variables $X$ to intensive ones $Y$, for instance.

Having introduced the effective relaxation time $\tau$, the notion of
{\it thermodynamic time} $\xi$
can now be extended for paths in more dimensions. Formally,
we retain the old definition (\ref{dxi}) of the  element of
thermodynamic time $\xi$:
\begin{equation}
d\xi=dt/\tau
\nonumber\end{equation}
which now depends on the direction of the speed $\dot X$.
Sometimes, a {\it vector of thermodynamic speed}
$X^\prime$ (or $Y^\prime$) will be more
useful than $\dot X$ (or $\dot Y$):
\begin{equation}
X^\prime\equiv {dX\over d\xi}=\tau\dot X~~~~or~~~
Y^\prime\equiv {dY\over d\xi}=\tau\dot Y.
\label{xprime}\end{equation}
The common (scalar) thermodynamic speed \cite{SalNul8588} corresponds
to the invariant norm(s) $\Vert X^\prime\Vert=\Vert Y^\prime\Vert$
of the vector(s) (\ref{xprime})
so an extension of the notion of thermodynamic speed for more dimensions
has been performed.
The entropy production $S^\prime=\tau\dot S$ per unit thermodynamic time
can also be considered. From Eq.~(\ref{tau3}) we obtain:
\begin{equation}
S^\prime\equiv\tau\dot S=\Vert X^\prime\Vert^2
\label{Sprime}\end{equation}
This means that the dimensionless entropy production rate
$S^\prime$ is equal to the squared thermodynamic speed,
as is expected from the single variable case in Sect.~2.

\section{Optimum paths minimizing entropy production}

Consider a certain path $\{Y(t)\}$ corresponding to a finite-time
thermodynamic process connecting the initial state $Y_i\equiv Y(t_i)$
with the final one $Y_f\equiv Y(t_f)$. Remember that a thermodynamic path
$\{Y(t)\}$ is the solution to the "cooling" equation (\ref{coolingeq})
having now the following general form [cf. Eq.~(\ref{dotxy1})]:
\begin{equation}
\dot Y=g\gamma \left( Y_r-Y \right),
\end{equation}
driven by the given reservoir path $\{Y_r(t)\}$. The system's path has a
small retardation $\Delta Y(t)$ behind the reservoir path.
In the slow process approximation Eq.~(\ref{dots1}) applies
and the total entropy production will depend on the path as follows:
\begin{equation}
S_{fi}\equiv\int_{t_i}^{t_f}\dot S dt
      =\int_{t_i}^{t_f}(\dot Y\vert (g \gamma g)^{-1}\vert \dot Y)dt.
\label{sfi}\end{equation}
Let us find the path minimizing the overall entropy production, among
paths connecting the fixed initial and ending points at a fixed time
lapse $t_f-t_i$. An analogy with Lagrange's variational principle can
be established if we identify the Lagrange-function as $\dot S/2$.
Then, minimizing paths are found to obey the following Euler--Lagrange--
equations:
\begin{equation}
{d\over dt}\left( (g \gamma g)^{-1}\dot Y\right)^k={1\over2}
(\dot Y\vert {\partial(g\gamma g)^{-1}\over\partial Y_k}\vert \dot Y).
\label{Eulerx}\end{equation}
Obviously, the entropy production rate (\ref{dots1}) is an integral of
the above differential equation:
\begin{equation}
\dot S = (\dot Y\vert (g \gamma g)^{-1}       \vert \dot Y) =const.
\label{sconst}\end{equation}
which is the mathematical counterpart of the energy conservation rule
in mechanics. In our case this means that
the entropy minimizing path corresponds to a constant entropy
production rate, as was pointed out in Sect.~2 for the single variable
cooling process.

An equivalent Euler equation could be obtained had we chosen the extensive
variables $X$ to parametrize the paths. Actually we have chosen the
intensive ones since in typical experimental situations the reservoir's
intensive variables are under our control (cf. Sect.~2).

In certain cases (see Sect.~5)
it would be interesting to find the path minimizing
$S_{fi}$ at the condition that the {\it thermodynamic time}
$\xi_f-\xi_i=\int_{t_i}^{t_f}dt/\tau$
be kept fixed and the clock-time $t_f-t_i$ might be varied. In this case,
one suitably replaces all $t$-dependences by $\xi$-dependences.
{}From rate (\ref{Sprime}), one obtains the
following equation for the total entropy production:
\begin{equation}
S_{fi}\equiv\int_{\xi_i}^{\xi_f} S^\prime d\xi
	=\int_{\xi_i}^{\xi_f}\Vert Y^\prime\Vert^2d\xi.
\end{equation}
At fixed $\xi_i,\xi_f$, the minimum of entropy production is achieved
if the path $Y(\xi)$ satisfies the following Euler-Lagrange equations:
\begin{equation}
{d\over d\xi}\left(g^{-1}Y^\prime\right)^k
={1\over2}(Y^\prime \vert {\partial g^{-1}\over\partial Y_k}\vert Y^\prime).
\label{Eulerxxi}\end{equation}
The entropy production rate versus thermodynamic time $\xi$ will be
an integral of this Euler-Lagrange equation:
\begin{equation}
S^\prime = \Vert Y^\prime\Vert^2 =const.
\label{sconstxi}\end{equation}
Thus the minimizing path corresponds to constant $S^\prime$. On the
other hand, $S^\prime$ is equal to the squared invariant norm
of the thermodynamic speed $Y^\prime$ defined by Eq.~(\ref{xprime}).
Hence the optimum path is of {\it constant thermodynamic speed}.
An equivalent result could be obtained in $X$-variables. The constancy of
the thermodynamic speed is merely a consequence of a much remarkable
feature of optimum paths: they are {\it geodesics} of the
Riemann--metricized manifold of thermodynamic states.

\section{An easy control of optimum cooling}

It follows from the previous Section that the Riemannian geometric structure
of the thermodynamic state space has an intrinsic relation with the
processes of maximum reversibility at fixed {\it thermodynamic} lapse rather
than clock-time. We are going to show that optimum cooling processes can
easily be controlled by using thermometers instead of clocks! The method is
straightforward to implement for the simultaneous control of more
thermodynamic parameters.

We propose a stepwise control strategy to approximate the optimum process
at constant thermodynamic speed. Assume that we change the reservoir
temperature stepwise between $T_i$ and $T_f$:
$$
T_i\equiv T_0=T_{r0}>T_{r1}>T_{r2}>\dots>T_{rN}=T_f,
$$
and the cooling strategy goes like this. First we lower the reservoir
temperature to $T_{r1}$ suddenly, and we wait for the system to become
equilibrated to a given extent $\epsilon$, i.e. its temperature $T$
will be as close to $T_{r1}$ as $T_1=\epsilon T_{r1}+(1-\epsilon)T_i$.
Then we move to the next iteration by lowering the reservoir temperature to
$T_{r2}$ and letting the system's temperature $T$ to equilibrate to the
{\it same} extent $\epsilon$,
identified by $T_2=\epsilon T_{r2}+(1-\epsilon)T_1$. In general:
\begin{equation}
T_{n+1}=\epsilon T_{r,n+1}+(1-\epsilon)T_n,~~~~(n=0,1,\dots,N-1).
\label{iter}\end{equation}
If all steps are so small that the change of the relaxation time $\tau(T)$
is negligible at the step's scale then the relaxation equation
(\ref{coolingeq}) holds and each step will have the {\it same} lapse
\begin{equation}
\delta\xi=\log(1-\epsilon)^{-1}
\label{deltaxi}\end{equation}
of thermodynamic time. Hence, the number $N$ of small steps required to
realize the cooling process from $T_i$ to $T_f$, provided the quality of
each relaxation has had the common characteristic value $\epsilon$, will
be proportional to the thermodynamic time lapse $\xi_f-\xi_i$ of the
cooling path:
\begin{equation}
N={\xi_f-\xi_i\over\vert\log(1-\epsilon)\vert}.
\label{N}\end{equation}
The smaller the defect of relaxations $1-\epsilon$ the bigger number of
iterations will be necessary to achieve the same cooling
$T_i\rightarrow T_f$.

To assure constant thermodynamic speed in average we choose steps
$T_{r1},T_{r2},\dots$ in such a way that the {\it same} thermodynamic length
$\delta\ell$ defined by (\ref{dTnorm2}) belong to all corresponding segments
$\Vert T_i-T_1\Vert,~\Vert T_1-T_2\Vert,\dots$:
\begin{equation}
{C(T_n)\over T_n^2}\left\vert T_{n+1}-T_n\right\vert^2=\delta\ell^2,~~~~
						(n=0,1,\dots,N-1)
\label{deltaell}\end{equation}
{}From equations (\ref{iter},\ref{deltaell}) one obtains:
\begin{equation}
T_{r,n+1}
=\left(1-{\delta\xi/\epsilon\over\sqrt{C(T_n)}}\right)T_n,~~~~
						(n=0,1,\dots,N-1).
\label{Trnplus1}\end{equation}

Now we can summarize the iterative algorithm of cooling.
Given the current temperature $T_n$ of the system,
one decreases the reservoir's
temperature according  to Eq.~(\ref{Trnplus1}) and lets the system
relax until the condition (\ref{iter}) becomes valid.
This procedure is then iterated for $T_{n+1}$, e.t.c. One can see by
inspection that the same average thermodynamic speed
\begin{equation}
v={\delta\ell\over\vert\log(1-\epsilon)\vert}
\label{t_speed}\end{equation}
belongs to each steps and consequently to the whole process as well.
So this stepwise algorithm approaches the theoretical optimum process
of constant thermodynamic speed \cite{foot1}.
We need no clocks to measure time but thermometers to measure temperatures
and the {\it a priori} knowledge of the specific heat function $C(T)$ of the
system in hand.

In the general case
when more thermodynamic variables are to be controlled one has to
derive the geodesic path between the initial and final equilibrium
states in advance and then apply the stepwise algorithm along
the geodesic path. It is worth noting that in simulated annealing
\cite{SalNul8588} the above algorithm can not be directly applied since the
annealed system's temperature and specific heat are  known but
statistically. We nevertheless think that our results could be adapted
to computer simulated statistical systems.

A few words are needed to interprete the condition of fixed thermodynamic
time when looking for the optimum control. It may often happen that
it is {\it not} the clock-time of a given thermodynamic process that is the
economically or technologically relevant quantity. Rather than clock-time,
the {\it number} of iterated technological steps might better characterize
the amount of relevant facilities (computer capacity, special materials, or
just money) that can be consumed to bring the system from its initial state
into a prescribed final one. In such cases the clock-time period of the
process is of less interest to be fixed in advance. The step number is
rather to be fixed.
The condition of fixed-in-advance step number
is equivalent to the condition of fixed-in-advance thermodynamic time, as
shown by Eq.~(\ref{N}).
It may eventually happen that the thermodynamic time (proportional to the
step number) takes the place of the clock-time.
The overall clock-time needed to perform a single cooling
step might not be dominated by the clock-time of the relaxation but
by the clock-time of the technological adjustment before and after the
relaxation. In this case, the gross time of the stepwise process
will be proportional to its thermodynamic time. Actually, the
fixed-in-advance thermodynamic time becomes the relevant condition in
designing the finite-time thermodynamic process and, consequently,
minimum entropy production will be achieved along the
geodesic path in thermodynamic state space.

\section{Conclusion}

We have extended the notion of the relaxation time $\tau$ from one dimension
to an arbitrary number of dimensions. Thus the thermodynamic time can be
deduced for multi-dimensional thermodynamic processes, too, in a
straightforward way. As a consequence, the optimum control of a
thermodynamic process based on the thermodynamic time scale can be achieved
with any number of variables to be controlled.
We have seen that the optimum path should be followed as close
as possible by the process in hand as the geodesic is uniquely marked out
between the initial and the desired final states. In addition, the proposed
stepwise driving force for the process can be arranged by measuring ordinary
thermodynamic variables e.g. temperature, concentration etc.

A remarkable result of the present analysis is that the path of minimum
irreversible entropy production in fixed thermodynamic time becomes a
geodesic of the Riemannian space of thermodynamic states introduced a
decade ago. Thus far Riemannian structure was seen
only in its influence in infinitesimal neighbourhoods,
e.g., on thermodynamic fluctuations. Now we have
shown within the context of {\it standard} irreversible thermodynamics that
some finite time thermodynamic processes, when optimum, follow the shortest
paths of the Riemannian space. The result is valid in an arbitrary number of
dimensions and proves for the first time the distinguished role of
geodesics in driven irreversible processes. Our result is based on the usual
expansion of the irreversible entropy production
up to the second order in speeds. Of course, the given interpretation for
geodesics is restricted to not too high speeds and rates. (This, however,
would not at all restrict the ultimate value of the interpretation.
Recall the analogous case in space-time geometry: it is not the too large
objects but the small enough ones that will follow geodesics.)
These results may turn out to form a basis for amplifying
the benefits of finite-time thermodynamics both in theory and in practice.

After the completion of our work
we became aware of a related report by Spirkl and Ries \cite{SpiRie95}.
Whereas they do not use Riemann-geometrical terms
there seem to be significant coincidences between our results.

\section{Acknowledgements}

This work is supported by the {\it Comission of European Communities} in the
frame-work of the "Copernicus" program and by the OTKA under Grant
No. T/007410.


\begin{thebibliography}{99}
\bibitem{AndSalBer84} B.Andresen, P.Salamon, and R.S.Berry,
	Physics Today {\bf 37}, 62 (1984) and references therein.
\bibitem{Wei} F.Weinhold, Phys. Today {\bf 29}, 23 (1976) and references
	therein.
\bibitem{Rup} G.Ruppeiner, Phys.Rev. {\bf A20}, 1608 (1979).
\bibitem{Dio} L.Di\'osi, G.Forg\'acs, B.Luk\'acs, and H.L.Frisch,
	Phys.Rev. {\bf A29}, 3343 (1984).
\bibitem{SalNul8588} P.Salamon, J.Nulton, and R.S.Berry, J.Chem.Phys.
	{\bf 82}, 2433 (1985); P.Salamon {\it et al.}, Comput.Phys.Commun.
	49, 423 (1988); J.Nulton and P.Salamon, Phys.Rev. {\bf A37}, 1351
	(1988); B.Andresen and J.M.Gordon, Open Systems \& Information
	Theory {\bf 2}, 1 (1993).
\bibitem{SalBer83} P.Salamon and R.S. Berry, Phys. Rev. Lett. {\bf 51}, 1127
	(1983).
\bibitem{AndGor94} B.Andresen and J.M.Gordon, Phys. Rev. {\bf E50},
	4346 (1994).
\bibitem{Lan} L.D.Landau and E.M.Lifshitz, {\it Statistical Physics},
	Clarendon, Oxford (1982).
\bibitem{foot1}
This stepwise control strategy leads exactly to the
continuous process of constant thermodynamic speed in the limit
when, at fixed $\xi_f-\xi_i$,
the number $N$ of steps goes to infinity while the step size
$\delta\xi=(\xi_f-\xi_i)/N$ goes to zero. In this limit,
Eq.~(\ref{deltaxi}) yields $\epsilon=\delta\xi$.
Detailed calculations, to be published elsewhere, lead to the
following asymptotic behaviour of the entropy production in the
stepwise process versus that in the continuous process:
$S_{fi}(step)
=\left(1+\ell C^{-1/2}_{av}/N+{\cal O}(1/N^2)\right)S_{fi}(cont)$,
where $\ell$ is the thermodynamic length of the continuous path and
$C^{-1/2}_{av}$ is the average of $C^{-1/2}(T_n)$ over
$n=1,2,\dots,N$.
\bibitem{SpiRie95} W.Spirkl and H.Ries, Munich report (1995).

\end{thebibliography}
\end{document}